\documentclass{pas}

\usepackage{natbib}
\usepackage[nolist]{acronym}

\usepackage{multirow}

\begin{acronym}
\acro{LMXB}{low mass X-ray binary}
\acro{NS}{neutron star}
\acro{MCMC}{Markov-Chain Monte Carlo}
\acro{AMP}{accreting millisecond pulsar}
\acro{RMS}{root mean squared}
\end{acronym}

\newcommand{\sigr}{IGR~J17511$-$3057}
\newcommand{\ssax}{SAX~J1808.4$-$3658}

\newcommand{\swiftscaling}{$1.3229\pm0.0053$}
\newcommand{\numbursts}{24}
\newcommand{\numburstsunique}{21}
\newcommand{\numburstsbeans}{20}
\newcommand{\xlo}{0.03} 

\newcommand{\eps}{{\rm erg\,s^{-1}}}
\newcommand{\epcs}{{\rm erg\,cm^{-2}\,s^{-1}}}

\newcommand{\mns}{\ensuremath{M_{\rm NS}}}
\newcommand{\rns}{\ensuremath{R_{\rm NS}}}

\newcommand{\zcno}{\ensuremath{Z_{\rm CNO}}}
\newcommand{\qb}{\ensuremath{Q_b}}
\newcommand{\xte}{{\it RXTE}}

\newcommand{\igr}{{\it INTEGRAL}}
\newcommand{\msun}{M_\odot}

\usepackage{multirow}

\begin{document}

\lefttitle{Publications of the Astronomical Society of Australia}
\righttitle{Cambridge Author}

\jnlPage{1}{13}
\jnlDoiYr{2026}
\doival{10.1017/pasa.xxxx.xx}

\articletitt{Research Paper}

\title{A helium-rich mass-donor in a 3.5-hr low mass X-ray binary}

\author{\sn{Galloway} \gn{Duncan K.}$^{1,2}$,
\sn{Hirai} \gn{Ryosuke}$^{3,1,4}$ and 
\sn{Goodwin} \gn{Adelle J.}$^{5}$}
\affil{$^1$School of Physics and Astronomy, Monash University, Victoria 3800, Australia; 
$^2$Institute for Globally Distributed Open Research and Education (IGDORE); 
$^3$Astrophysical Big Bang Laboratory (ABBL), Pioneering Research Institute (PRI), RIKEN, Wako, Saitama 351-0198, Japan; 
$^4$OzGrav: The ARC Centre of Excellence for Gravitational Wave Discovery, Clayton, VIC 3800, Australia; 
$^5$International Centre for Radio Astronomy Research – Curtin University, GPO Box U1987, Perth, WA 6845, Australia}

\corresp{D.~K. Galloway, Email: duncan.galloway@monash.edu}

\citeauth{Galloway, D.~K, Hirai, R. and Goodwin, A.~J, A helium-rich mass-donor in a 3.5-hr low mass X-ray binary. {\it Publications of the Astronomical Society of Australia} {\bf 00}, 1--13. https://doi.org/10.1017/pasa.xxxx.xx}

\history{(Received xx xx xxxx; revised xx xx xxxx; accepted xx xx xxxx)}

\begin{abstract}
Binary systems in which a neutron star accretes material from a low-mass stellar companion arise from evolutionary processes that depend sensitively on the progenitor properties.
It is usually assumed that the accreted material has approximately solar composition, unless the orbital period is $\lesssim 80$~min, indicating an ``ultracompact'' system with a hydrogen deficient donor. However, few constraints on the donor composition are available. Here we apply a Bayesian model-comparison framework, {\sc beansp}, to a sample of \numburstsunique\ thermonuclear bursts from the accretion-powered millisecond pulsar, IGR~J17511$-$3057, to constrain the fuel composition as well as the system parameters. Our analysis strongly indicates an extremely hydrogen- and CNO-deficient fuel composition, with $X\approx0.03$ and $\zcno\approx4\times10^{-4}$, unexpected for this system with its relatively long 3.5~hr orbital period. 
We explore the possible circumstances under which such an object can evolve, and discuss the implications for its history also in light of the three transient outbursts that have been observed to date.
Such analyses are finding an unexpectedly wide range in donor compositions, that may help to explain the remarkable diversity in the burst properties of the known burst sources.
\end{abstract}


\maketitle

\section{Introduction}
\label{sec:intro}

\Acp{LMXB} consisting of a neutron star accreting from a main-sequence or evolved donor are identifiable thanks to their bright X-ray emission during 
accretion activity 
\cite[]{baglio26}.
Such objects are the exceedingly rare outcome of a complex interplay between stellar and binary evolution \cite[e.g.][]{prp02,prp03}, and only a few hundred examples are known.

The principle diagnostic of the evolutionary history, beyond the orbital period, is the donor composition, which can be measured via spectroscopy \cite[e.g.][]{cs24}. However, such objects also tend to be distributed close to the Galactic centre, where extinction from dust and gas make the optical counterparts difficult to observe.
An alternative way of constraining the donor composition is by analysing the fuel composition for thermonuclear (type-I) X-ray bursts, which can be distinguished from the persistent X-ray emission arising from accretion given X-ray detectors with sufficient sensitivity and time resolution. Such events arise from unstable ignition of the accumulated fuel on the surface of the \ac{NS} 
and exhibit properties that depend on the fuel composition at ignition
\cite[e.g.][]{gal21a}.

The difficulties in constraining the accreted fuel composition from the bursts are threefold; first, the composition at ignition is modified by prior steady burning \cite[most notably at high accretion rates; e.g.][]{cavecchi26}; second, the observable properties of individual bursts (duration, total energy, recurrence time) are only weakly sensitive to the accreted composition; and third, the burst and persistent emission properties are modified by the confounding effects of the host system, including the accretion disk.
To overcome these difficulties we have developed the {\sc beansp} code which compares observed burst properties to the predictions of a numerical ignition model {\sc settle} \citep[based on the ignition calculations of][]{cumming03}, including the properties of the host \cite[]{goodwin19c}. We use a \ac{MCMC} approach to fully explore the parameter space and establish the posterior distributions of our composition and system parameters. 

We have so far focused our analysis on transient accretion outbursts in \acp{AMP}, as these sources are even rarer than normal \acp{LMXB} and hence are high observational priority when they are active \cite[e.g.][]{disalvo22}, leading to a high probability of detecting bursts. Such objects also tend to exhibit H-deficient bursts, for which our ignition code is known to produce more reliable predictions \cite[]{gal06c}.

In this paper we attempt to match the properties of thermonuclear bursts from the \ac{AMP} \sigr\ to our numerical model, with a view to constraining the system properties including fuel composition.
\sigr\ consists of a neutron star rotating with a spin frequency of 244.8~Hz, and a binary companion orbiting once every 3.47~hr \cite[]{riggio11}. The source was discovered by \igr\ when it went into outburst in 2009 September \cite[]{falanga11}. Thermonuclear bursts were detected during the outburst, first by {\it Swift}\/ \cite[]{bozzo10a}, as well as \xte/PCA and \igr/JEM-X. The bursts were short duration (7--10~s), none with evidence for photospheric radius expansion, and exhibited burst oscillations at a frequency consistent with the persistent pulsations \cite[]{watts09a}. The bursts were consistent with recurrence times in the range 7--15~hr, which, along with the fluence measurements, leads to a limit on the fuel H-fraction of $X\leq0.44$ (for CNO metallicity $\zcno=0.02$). \cite{falanga11} also noted that the $\alpha$-value, the ratio of the persistent to burst luminosity, decreased with increasing burst rate, the opposite of the expected trend.
Via spectral fitting of the iron K$\alpha$ line, \citet{papitto10} constrained the system inclination to be in the range 38--68\,deg.
Since discovery, the source has been active twice more, in 2015 \cite[]{papitto16} and 2025 \cite[e.g.][]{illiano25}.

This paper is part of a long-running series (\citealt{goodwin19c,gal24,gal26} most recently, but beginning with \citealt{gal06c}) intended to establish the method for constraining the properties of bursting neutron stars, and as such should not be considered the final word on this approach. Rather, we continue to learn from the application of the method to new sources and to improve the procedure for more reliable performance.
In Section \ref{sec:obssim} we describe the observations and data preparation, and present a basic description of the simulation procedure.
In Section \ref{sec:results} we describe the outcomes of our analysis and an examination of the outburst history.
In Section \ref{sec:discussion} we consider the possible evolutionary scenarios that might have given rise to this system.

\section{Observations \& Simulations}
\label{sec:obssim}

Our primary analysis focussed on a \numburstsunique-burst sample assembled from the Multi-INstrument Burst ARchive, MINBAR \citep{minbar}, comprising bursts observed from \sigr\ during the 2009 outburst with \xte/PCA and \igr/JEM-X, along with bursts from {\it Swift} and {\it XMM-Newton} \cite[]{bozzo10a, papitto10, falanga11}. 
The details for the \numbursts\ observed events (\numburstsunique\ unique) assembled for the analysis are provided in Table \ref{tab:bursts}.

We  adopted estimates of the bolometric persistent flux during the 2009 outburst from \cite{falanga11}.
For the bursts observed during that outburst with JEM-X from MINBAR, we adopted the fluences listed in \cite{falanga11}. The full burst dataset is listed in Table \ref{tab:bursts}.
Because the persistent flux measurements did not extend sufficiently early compared to the time of burst \#1, we omitted that event from the {\sc beansp} simulations, leaving a comparison sample of \numburstsbeans\ events.

\begin{table*}
\caption{Bursts observed from \sigr\ during the 2009 outburst. 
} \label{tab:bursts}
 {\tablefont\begin{tabular}{@{\extracolsep{\fill}}ccccccccccl}
 \toprule
& \multicolumn2c{Burst start time}  &  &  & Sep. & \multicolumn2c{Burst fluence ($10^{-6}\ {\rm erg\,cm^{-2}}$)} & Peak flux & \\
No. & Calendar (UTC) & MJD (d) & Instr.$^{\rm a}$ & Dup.?$^{\rm b}$ & hr & original & adopted & ($10^{-9}\ \epcs$) & $\alpha$ & Ref.$^{\rm c}$ \\
 \hline
1 & 2009-09-12 04:03:31 & 55086.16911 & IJ1 & N & 0.00 & \ldots & \ldots & \ldots & \ldots & 7619 \\
2 & 2009-09-14 00:50:27 & 55088.03503 & {\it Swift} & N & 44.78 & $0.27\pm0.02$ & $0.36\pm0.03$ & $40\pm8$ & \ldots & [1] A; [2] 1 \\
3 & 2009-09-14 07:54:45 & 55088.32969 & XPp & N & 7.07 & $0.351\pm0.003$ & $0.351\pm0.003$ & $38.2\pm1.4$ & $123.5\pm1.0$ & 3858; [2] 2 \\
4 & 2009-09-15 03:57:35 & 55089.16499 & IJ1 & N & 20.05 & \ldots & \ldots & \ldots & \ldots & 7630 \\
5 & 2009-09-15 17:17:19 & 55089.72036 & {\it Swift} & Y & 13.33 & $0.28\pm0.02$ & $0.37\pm0.03$ & \ldots & \ldots & [1] B; [2] 3 \\
5 & 2009-09-15 17:17:25 & 55089.72043 & XP2 & N & 0.00 & $0.376\pm0.005$ & $0.376\pm0.005$ & $36.2\pm1.3$ & $112.2\pm1.4$ & 3859; [2] 3 \\
6 & 2009-09-17 06:33:28 & 55091.27324 & IJ1 & Y & 37.27 & \ldots & $0.374\pm0.004$ & \ldots & \ldots & 7653; [2] 4 \\
6 & 2009-09-17 06:33:29 & 55091.27325 & XP2 & N & 0.00 & $0.398\pm0.005$ & $0.398\pm0.005$ & $44.9\pm1.8$ & $91.7\pm1.2$ & 3860 \\
7 & 2009-09-17 14:48:43 & 55091.61717 & XP2 & N & 8.25 & $0.384\pm0.013$ & $0.384\pm0.013$ & $42.2\pm1.6$ & $124\pm4$ & 3861; [2] 5 \\
7 & 2009-09-17 14:48:49 & 55091.61723 & IJ1 & Y & 0.00 & \ldots & \ldots & \ldots & \ldots & 7656 \\
8 & 2009-09-18 08:08:26 & 55092.33919 & IJ1 & N & 17.33 & \ldots & $0.33\pm0.06$ & $45\pm9$ & $140\pm30$ & 7658; [2] 6g \\
9 & 2009-09-18 17:40:16 & 55092.7363 & IJ1 & N & 9.53 & \ldots & \ldots & \ldots & \ldots & 7676 \\
10 & 2009-09-19 02:27:18 & 55093.10229 & IJ1 & N & 8.78 & \ldots & $0.34\pm0.07$ & $38\pm9$ & $140\pm30$ & 7698; [2] 7g \\
11 & 2009-09-19 11:11:59 & 55093.46665 & IJ1 & N & 8.74 & \ldots & $0.39\pm0.07$ & $37\pm10$ & $110\pm20$ & 7707; [2] 8g \\
12 & 2009-09-20 14:50:33 & 55094.61844 & XPr & N & 27.64 & $0.403\pm0.003$ & $0.403\pm0.003$ & $47\pm2$ & $105.5\pm0.7$ & 3862; [2] 9 \\
13 & 2009-09-21 01:12:22 & 55095.05025 & {\it XMM-Newton} & N & 10.36 & $0.48\pm0.11$ & $0.48\pm0.11$ & $42\pm11$ & $100\pm20$ & [3] I; [2] 10 \\
14 & 2009-09-21 12:34:02 & 55095.52363 & {\it XMM-Newton} & N & 11.36 & $0.49\pm0.12$ & $0.49\pm0.12$ & $43\pm11$ & $100\pm30$ & [3] II; [2] 11 \\
15 & 2009-09-22 12:54:59 & 55096.53818 & {\it Chandra} & N & 24.35 & \ldots & \ldots & \ldots & \ldots & [2] 12 \\
16 & 2009-09-23 14:27:09 & 55097.60219 & XP2 & N & 25.54 & $0.418\pm0.005$ & $0.418\pm0.005$ & $53\pm3$ & $114.0\pm1.3$ & 3863; [2] 13 \\
17 & 2009-09-24 18:39:15 & 55098.77726 & XP2 & N & 28.20 & $0.37\pm0.03$ & $0.37\pm0.03$ & $49\pm3$ & $135\pm10$ & 3866; [2] 14 \\
18 & 2009-09-25 07:31:38 & 55099.31364 & XPz & N & 12.87 & $0.425\pm0.003$ & $0.425\pm0.003$ & $58\pm2$ & $102.0\pm0.6$ & 3867; [2] 15 \\
19 & 2009-09-26 15:11:24 & 55100.63292 & XP2 & N & 31.66 & $0.42\pm0.02$ & $0.42\pm0.02$ & $60\pm3$ & $119\pm6$ & 3868; [2] 16 \\
20 & 2009-09-27 06:57:24 & 55101.28986 & XPd & N & 15.77 & $0.417\pm0.017$ & $0.417\pm0.017$ & $62\pm3$ & $115\pm5$ & 3869; [2] 17 \\
21 & 2009-09-30 18:31:57 & 55104.77219 & {\it Swift} & N & 83.58 & $0.48\pm0.03$ & $0.64\pm0.04$ & \ldots & $111\pm7$ & [1] C; [2] 18
   \botrule
    \end{tabular}}
    \begin{tabnote}
    {$^{{\rm a}}$Instrument name or MINBAR label, latter with two-character prefix for instrument (IJ = \igr/JEM-X, XP = \xte/PCA) and the third character corresponding to the camera choice/combination. See  \cite{minbar} for full details.}\tnp
    {$^{{\rm b}}$Indicates if the burst is a duplicate (Y/N), i.e. observed by more than one instrument. For such duplicates we included the most precise measurements if two estimates of the fluence were available.}\tnp
    {$^{{\rm c}}$References: four-digit numbers are the burst IDs in MINBAR. Other labels refer to 1. \cite{bozzo10a}; 2. \cite{falanga11}; 3. \cite{papitto10}.}\tnp
    \end{tabnote}
\end{table*}

\begin{figure}[ht!]
    \begin{centering}
        \includegraphics[width=\linewidth]{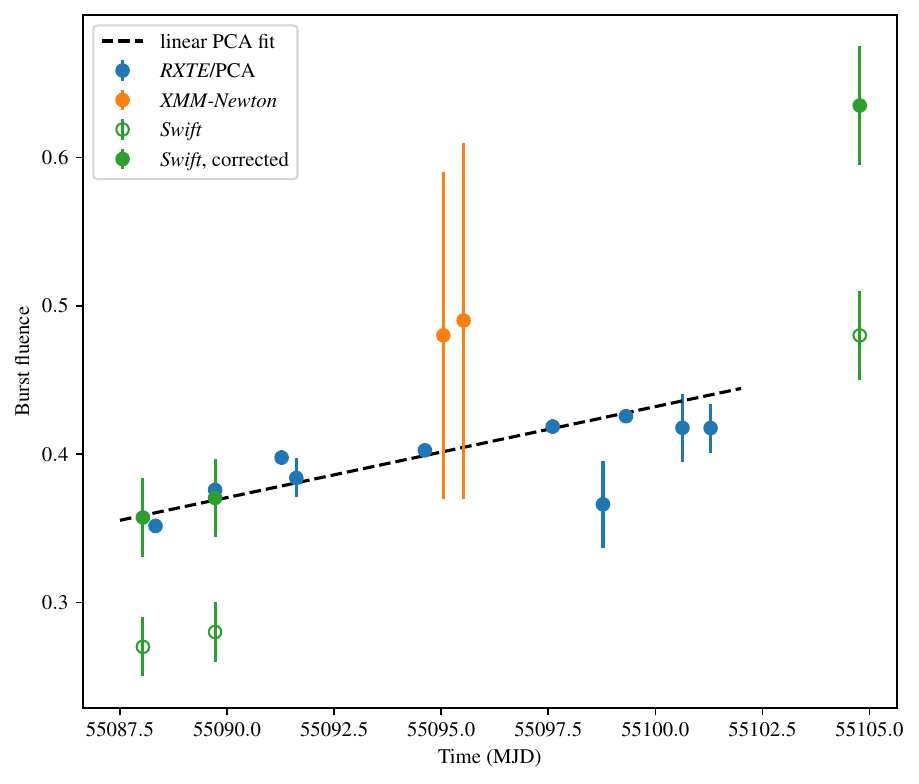}
        \caption{
            Variation in the fluence of bursts from \sigr\ as a function
of time, during the 2009 outburst. The dashed line is a linear fit to the
PCA data only; the closed green symbols labeled ``{\it Swift}, corrected''
represent the {\it Swift} measurements multiplied by the derived scaling
factor compared to PCA bursts, of  
\swiftscaling. }
        \label{fig:xcor}
    \end{centering}
\end{figure}

We identified discrepancies between the burst fluences measured with different instruments. The \xte/PCA bursts represent the largest single contribution (10 bursts), and featured fluences that increased steadily over the the 14 day period in which they were observed  (Figure \ref{fig:xcor}). The two bursts detected with {\it XMM-Newton}  had fluences that were substantially above the value measured for the nearest PCA burst, but with large uncertainties, so that the values were overall consistent with the PCA trend. Two of the three bursts detected by {\it Swift}\/ \cite[]{bozzo10a} occurred during the coverage with PCA (\#5 was observed with both instruments), and instead exhibited significantly {\it lower} fluences. We attribute the discrepancy to the relative flux cross-calibration of the two instruments. We corrected the {\it Swift} fluences by multiplying by a correction factor derived from the pair of bursts in the overlapping coverage; for burst \#2 
we adopted a fluence based on the overall trend line of the PCA data (Figure \ref{fig:xcor}), 
giving an overall ratio of
\swiftscaling.
Notably, this correction makes the third {\it Swift} burst (\#21) the brightest by far, and well in excess of the extrapolated trend line from the PCA observations.

\begin{figure}[ht!]
    \begin{centering}
        \includegraphics[width=\linewidth]{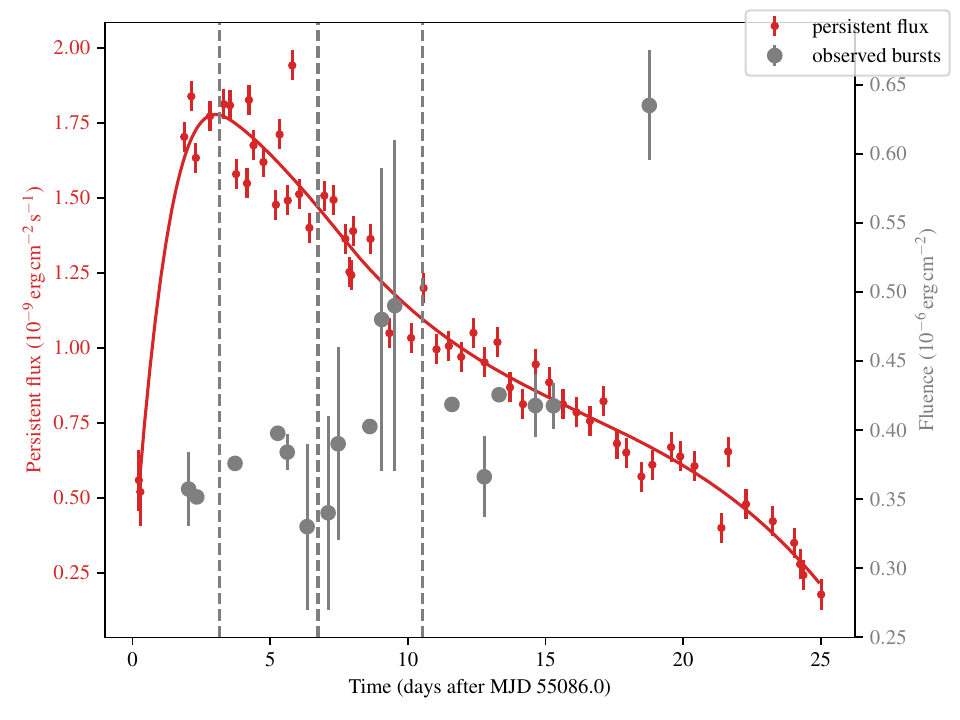}
        \caption{
            Bolometric persistent flux ({\it red symbols}, left-hand $y$-axis) of \sigr\ during the 2009 outburst, adopted from \cite{falanga11}. The solid red line is a spline fit to the data. The grey symbols are the corrected fluences (right-hand $y$-axis) for the burst sample; the times of bursts without a measured fluence are indicated by the gray dashed lines.
        }
        \label{fig:outburst}
    \end{centering}
\end{figure}

In addition to the burst fluence we incorporated measurements of the peak burst flux and the inferred $\alpha$-value into our analysis. We expect the burst peak fluxes to be limited by the Eddington flux appropriate for the source distance \cite[e.g.][]{kuul03a}; since none of the bursts exhibited photospheric radius-expansion, they should reach peak fluxes below this value.
The $\alpha$ value is the integrated ratio of the persistent flux to the burst flux, i.e.
\begin{equation}
    \alpha = \frac{\int_0^{\Delta t} F_X dt}{E_b}
    \label{eq:alpha}
\end{equation}
where $\Delta t$ is the burst recurrence time, $F_X$ is the persistent flux (which varies throughout the outburst), and $E_b$ is the burst fluence. This quantity measures the relative efficiency of accretion to nuclear burning, and in conservative burning scenarios we expect a range of 40--150 depending on the fuel composition \cite[e.g.][]{concord22}.
Measuring alpha during transient outbursts with partial coverage is confounded by missing bursts, which may prevent unambiguously measuring $\Delta t$, and hence the interval over which the flux should be averaged. We overcame this difficulty by using the spline flux model (see Section \ref{sec:sim}), and using a burst count that is partially informed by the results of our simulations. For example, our simulations strongly indicate that three bursts were missed between the last pair of observed bursts, so we estimated the $\alpha$-value for that burst based on an estimate of $\Delta t=t_{\rm sep}/4$, where $t_{\rm sep}$ is the observed time separation of the two events, and integrating the flux model over only the last quarter of the interval to obtain the numerator in equation \ref{eq:alpha}. Overall we found a distribution of $\alpha=114\pm14$ and no strong evidence for the decreasing trend with time suggested by \cite{falanga11}.
The full 2009 outburst dataset used in this paper, including the corrected fluences for the {\it Swift}\/ bursts (but neglecting burst \#1 and the duplicates in Table \ref{tab:bursts}) is plotted in Figure \ref{fig:outburst}.

We also used estimates of the persistent flux during the 2015 and 2025 outbursts
to estimate the long-term average X-ray intensity.
For the 
2015 March outburst, we used flux measurements from {\it Swift-XRT}, {\it XMM-Newton}, and \igr/JEM-X and ISGRI, as presented in \cite{papitto16}. Those authors calculated the ratio of fluxes in the 0.5--100 and 0.5-10~keV range as $2.27\pm0.17$, and we adopt this value as the bolometric correction for outbursts \#2 and 3.
For the 2025 February outburst, we used 0.5--10~keV flux measurements from \igr, {\it NICER}, {\it NuSTAR},  and {\it XMM-Newton}, as compiled by \cite{illiano25}. 
A summary of the outburst history is provided in Table \ref{tab:outbursts}.

\begin{table*}
	\centering
 \caption{Properties of the outbursts in \sigr.
        \label{tab:outbursts} }
 {\tablefont\begin{tabular}{@{\extracolsep{\fill}}lcccccl}
   \toprule
     &             & Duration & Separation & Fluence  & $\left<F_x\right>$  & \\ 
Date & $t_0$ (MJD) & (d)      & (yr)       & $10^{-3}\ {\rm erg\,cm^{-2}}$ & $10^{-10}\ {\rm erg\,cm^{-2}\,s^{-1}}$ & Ref. \\
\hline
2009 Sep & 55086 & 30 & \ldots & 2.157 & \ldots & \cite{bozzo10a} \\
2015 Mar & 57105 & 29 & 5.53 & $1.94\pm0.15$ & $0.111\pm0.008$ & \cite{papitto16} \\
2025 Feb & 60717 & 22 & 9.89 & $0.98\pm0.08$ & $0.031\pm0.002$ & \cite{illiano25}
   \botrule
    \end{tabular}}
\end{table*}

\subsection{Burst simulations}
\label{sec:sim}

We used the {\sc beansp}\footnote{\url{https://github.com/outs1der/beans}} package to perform the model-observation comparisons, following \cite{gal24,gal26}.
The code performs a procedure analogous to radio pulsar timing, but using a numerical burst ignition model, and also compares the predicted fluences of the bursts as well as their onset times.

In its default mode of operation, {\sc beansp} generates a train of bursts using the {\sc settle} ignition model \cite[]{cum02} based on an inferred accretion rate history, derived in turn from a set of persistent flux measurements covering the interval of interest. The input parameters to the model are the fuel composition, H-fraction $X_0$ and CNO mass fraction \zcno; the ``base flux'' \qb, quantifying heat contributed from below the simulation region; the source distance $d$; anisotropy parameters $\xi_b$ and $\xi_p$ for the burst and persistent emission respectively, following the approach of \cite{he16}; and (optionally) the neutron star mass \mns\ and radius \rns. Each observed burst is matched to a simulated event, and an overall likelihood is calculated based on the agreement of the burst onset time; fluence $E_b$ and (optionally) the $\alpha$ and the peak burst flux. This algorithm is implemented at each time step for a set of Monte-Carlo ``walkers'', and the entire ensemble is iterated via the affine-invariant Markov-chain sampler {\sc emcee} \cite[]{emcee13}. The Python code has been developed over a series of papers beginning with \cite{goodwin19c}, and most recently \cite{gal24,gal26}; the algorithm was originally implemented in IDL and applied to \ssax\ \cite[]{gal06c}.

The dataset for \sigr\ presented the most significant challenge to date for the {\sc beansp} code, due to the large number (20) of bursts to match, as well as the extent of the data. A large number of observed bursts means a longer duration for each model realisation, but there is also the question of maintaining the burst count over the entire train. 
The total timespan of more than 18 days covered by the bursts, might be expected to require a simulated train of up to 60 bursts (assuming a steady recurrence time consistent with the shortest separation; Table \ref{tab:bursts}). 
For this reason, the path to the final parameter constraints was, frankly, torturous.

We attribute the difficulty in 
achieving acceptable simulation results
to two principal effects.
First, our knowledge of the accretion rate (derived from the persistent flux) unavoidably involves errors; because the original {\sc beansp} algorithm generates continuous burst chains, the cumulative effect of these errors is catastrophic for large burst samples (i.e. long burst trains), making matching them with observations increasingly challenging far from the reference burst (simulation start point). 

Second, simulating over long gaps 
between observed bursts introduces possible discontinuities in the likelihood hypersurface. 
We 
illustrate with 
the following possible scenario.
A particular region of parameter space will produce burst trains for which a specific subset will match with the observed bursts. The predicted trains must bridge long gaps between observed bursts (e.g. the 37.3-hr gap between 
bursts \#5 and 6, observed on MJD~55089.720 and 55091.273) with $n$ unseen bursts, required for the consistency of the model across the gap. As the walkers move into another region of parameter space, the average burst rate over the gap might increase or decrease slightly, such that a {\it different} subset of predicted bursts now provides the best match to the observed times. As a result, the best estimate of the number of unseen bursts in the gap is now $n\pm1$.
We expect such abrupt changes in the model to also result in step-changes in the model likelihood. Such steps may present a significant challenge for the MCMC algorithm in terms of finding the lowest global likelihood. We suggest that the strongly multi-modal posterior distributions from the MCMC runs, a common feature of the simulations carried out with this approach, may arise in part from these step-changes.

\begin{figure}[ht!]
    \begin{centering}
        \includegraphics[width=\linewidth]{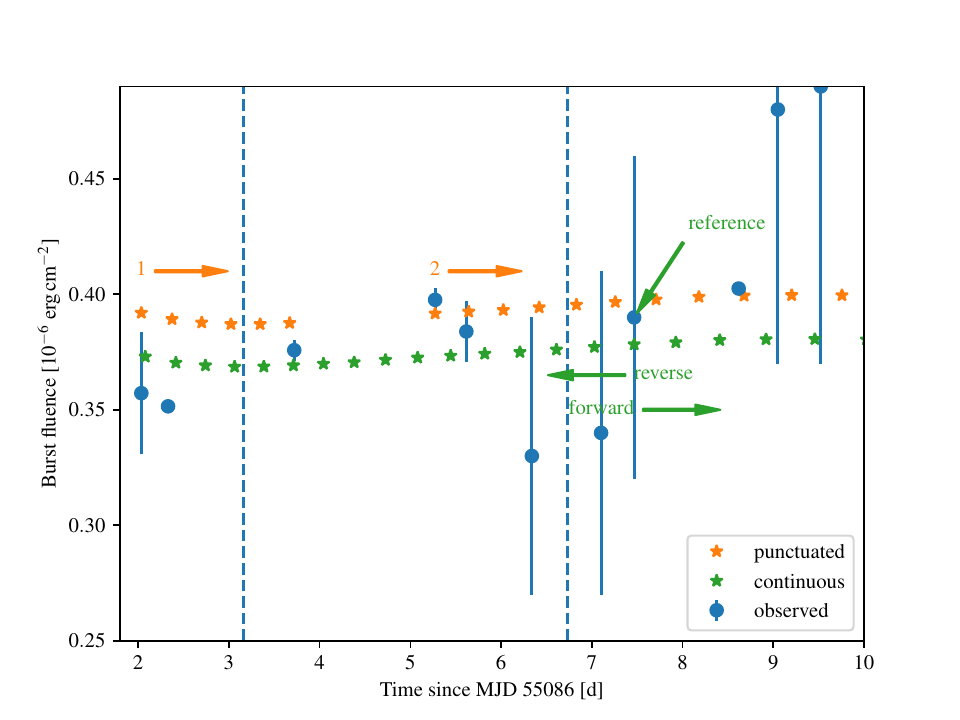}
        \caption{
            Example simulated bursts illustrating the different approaches available with the {\sc beansp} code. The observed fluences of bursts from \sigr\ are shown as blue symbols; bursts without measured fluences are marked as vertical dashed lines. An example of a continuous burst train, which is assembled by simulating in the forward and reverse directions simultaneously from the user-labeled reference burst,  is shown as green symbols. A piecewise continuous (``punctuated'') train is simulated in two parts, separated by the long gap between the 4th and 5th bursts, and is shown as the orange symbols. The interruption is necessitated by the excess number of bursts simulated in the gap, more than the user-specified value (as the {\tt maxgap} parameter).
        }
        \label{fig:example}
    \end{centering}
\end{figure}

As a result, we developed and implemented several improvements to the simulation and matching algorithm, as follows. First, we introduced the possibility of gaps in the predicted sequences. We begin with the first observed burst and generate a burst train moving forward in time, towards the next observed burst. Once the train goes beyond the next observed burst, we match with the predicted event that occurs closest to the time of that burst. However, if the interval requires an excessive number of missed bursts 
(we used a limit of 3), we omit the comparison between the following observed and predicted bursts from the likelihood calculation, and {\it restart} the train with the next observed burst. 
The difference between this approach and that used previously is illustrated in Fig. \ref{fig:example}.

Second, to avoid the errors that accumulate in the predicted burst sequences due to imperfect knowledge of the accretion rate, we 
simulate each burst forward from the {\it previous observed burst}, instead of the previous {\it simulated} event. In this manner the prediction for the $i$th observed burst is based 
only on the time of the $(i-1)$th observed burst and the persistent flux history since,
and hence is independent of the 
simulation history before (or after) that event.
And third, we introduce an additional multiplicative parameter $f_t$ to account for the systematic variations in the burst times, that may be present even despite the first two modifications.

\subsection{Stellar simulations}
\label{sec:mesa}

To 
establish the possible evolutionary history of \sigr, we constructed stripped post-main-sequence stellar models with the public stellar evolution code MESA \citep[v24.08.01][]{MESA1,MESA2,MESA3,MESA4,MESA5,MESA6}. 
As we argue in Section \ref{sec:donor}, the most likely route to H-deficient mass donors is a post-main sequence star stripped down to the layer immediately outside the helium core, by an interval of unstable mass transfer culminating in a common-envelope phase.

We thus evolve stars with zero-age main-sequence masses in the range $3\leq M_\mathrm{ZAMS}/\msun\leq8$ until their radii reach three times their terminal-age main-sequence values. At this stage, we assume that the star undergoes a common-envelope event and removed its envelope down to the mass coordinate 
corresponding to the estimated H-composition established from {\sc beansp},
using the \texttt{relax\_mass} feature in {\sc mesa}. The stripped models are then evolved until core helium exhaustion, and we examine how the radius of such objects evolve with a view towards verifying the conditions for Roche-lobe overflow with the observed orbital properties.

\section{Results}
\label{sec:results}

\begin{figure}[ht!]
    \begin{centering}
        \includegraphics[width=\columnwidth]{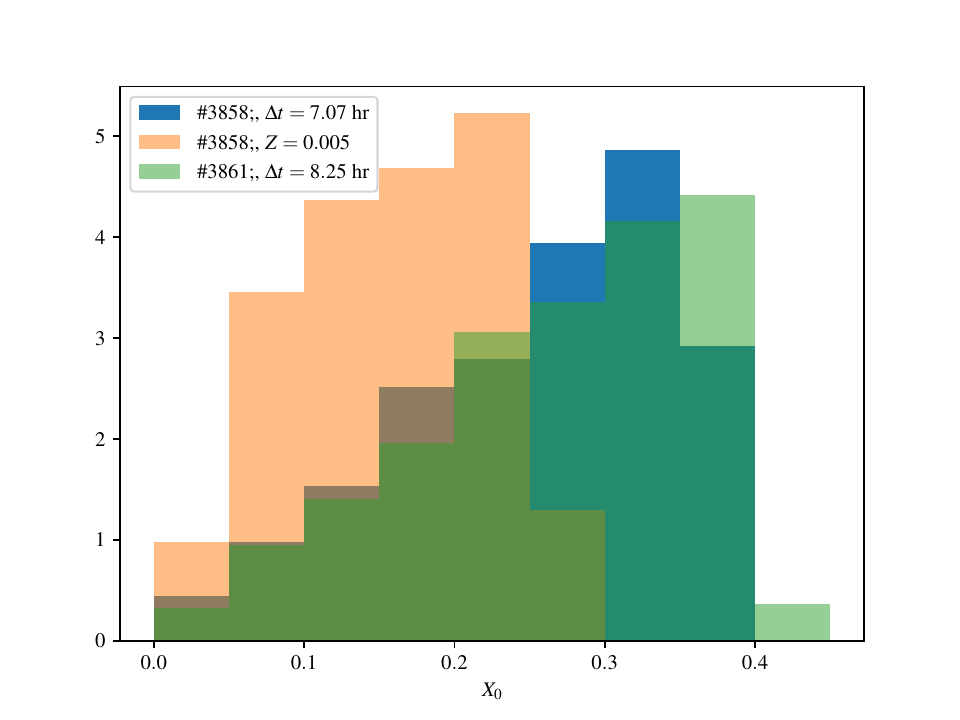}
        \caption{
            Inferred probability distribution function of the fuel H-fraction, $X_0$, based on the two shortest intervals between bursts. Two histograms are shown for burst \#3858, to indicate the effect of lower CNO metallicity.
        }
        \label{fig:X0}
    \end{centering}
\end{figure}

As a first attempt at constraining the burst fuel composition prior to any modelling, we  selected the two pairs of bursts with the shortest separation times, MINBAR \#3858 ($\Delta t=7.07$~hr) and \#3861 ($\Delta t = 8.25$~hr) 
We adopted the likely range of inclinations established by \cite{papitto10} of $i=38–68^\circ$, and assumed an isotropic distribution within this range. Preliminary analysis with {\tt concord} \citep{concord22} suggests sub-solar hydrogen fraction for the burst fuel, $X_0<0.4$, unless the metallicity is super-solar (Figure \ref{fig:X0}).
On the other hand, the additional distribution plotted for burst \#3858 but with $\zcno=0.005$ demonstrates that {\it sub-solar} metallicity implies an even smaller maximum value of $X_0$.

\subsection{Punctuated burst train simulations}
\label{sec:punkt}

We made a series of {\sc beansp} simulations 
motivating the ongoing code development described in Section \ref{sec:sim}, 
with the goal of replicating the observed burst behaviour of \sigr\ during the 2009 outburst.
We describe below our ``baseline'' model results, labeled {\tt pa\_mr} throughout.
We chose our initial walker positions for the runs presented here based on this series of tests and exploratory simulations,
with notably low input values for both the compositional parameters: 
$X\approx0.2$, $\zcno\approx0.003$.

We  ran the chains with 100 walkers typically for $3\times10^4$ steps or more. As is common with this approach, the walkers tended to split into groups in parameter space and remain nearby, even when the predicted burst properties in those regions replicated the observations poorly. For that reason, we refined the walker positions using the ``prune'' function of {\sc beans}, following \cite{gal26}. 
We reduced the ``stretch'' parameter of \cite{gw10} in steps from an initial value of 2 to ${\tt stretch\_a}=1.25$, to maintain the acceptance fraction within the range 0.2--0.5. 

As with previous applications of this algorithm, our runs do not achieve the criterion for convergence suggested by \cite{emcee13}. Instead, we ran the chains fur sufficiently long that the plotted walker ``chains'' (i.e. the trace plot) appear to reach a quasi-stationary state. As we hypothesize in our earlier paper, the lack of formal convergence may be a side-effect of the algorithm (rather than an indication of insufficient run length; \citep{gal26}). 
In addition to verifying qualitative convergence via the trace plots, we also ensure that the predicted model realisations for the posterior distributions replicate the observations.

\begin{figure*}[ht!]
    \begin{centering}
        \includegraphics[width=\linewidth]{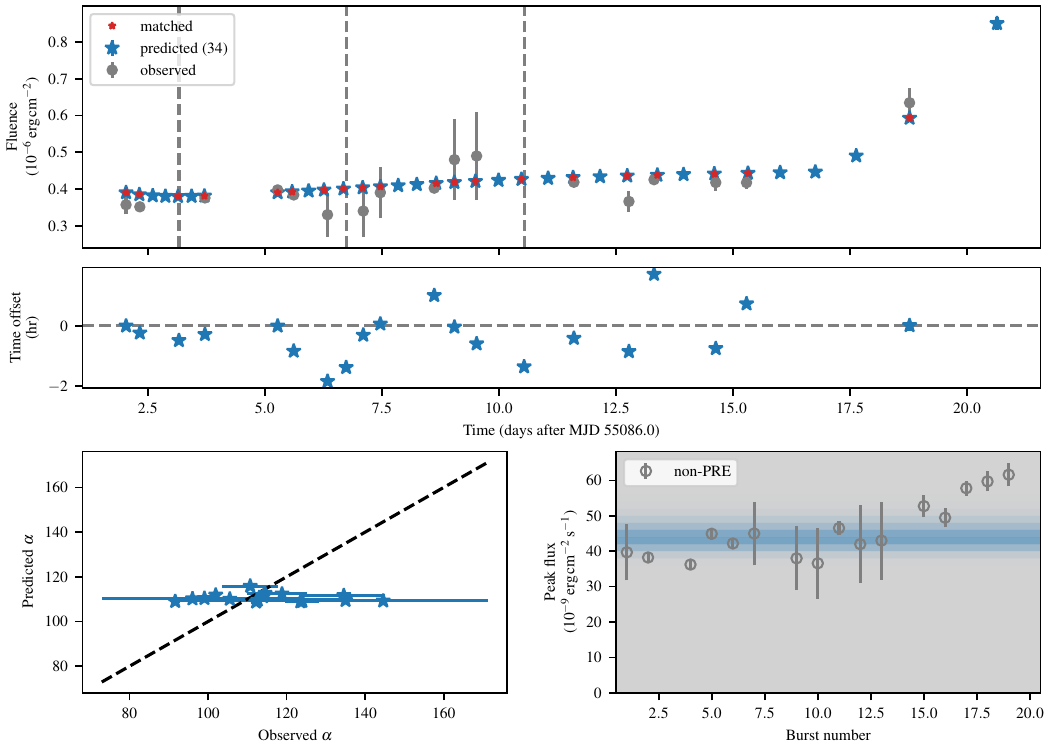}
        \caption{
            Comparison of the predicted and observed bursts from the 2009 outburst of \sigr, for the piecewise continuous (``punctuated'') independent burst train simulation labelled {\tt pa\_mr}.
            The top panel shows the observed burst fluence as a function of time ({\it gray symbols}); bursts where the fluence could not be measured are indicated as dashed gray lines. The ensemble prediction is shown as the blue stars, and the predictions additionally marked red are those predicted events that have been matched with the observed events.
            Note the substantial gap between the fourth and fifth observed bursts, and the last pair of observed bursts 
            between which the burst train was restarted.
            The middle panel shows the time residuals between the matched model-predicted burst times, and the observed burst times; the RMS error is 0.8535~hr.
            The bottom left-hand panel shows the predicted $\alpha$-values compared to the observed estimates.
            The right hand panel compares the peak flux of the bursts, with the distribution of expected Eddington fluxes for the inferred source distance shown as the blue bands. This panel is greyed out to indicate the peak fluxes were not considered in the likelihood.  
        }
        \label{fig:punkt_ind_mrt_pred}
    \end{centering}
\end{figure*}

The fluences were reproduced reasonably well (Figure \ref{fig:punkt_ind_mrt_pred}), with roughly constant values (as observed) up to about 15~d after the outburst start, after which the fluence for the predicted bursts increases more steeply. This increase has the effect of also matching approximately the fluence of the last (and most energetic) observed burst, \#21.

The predicted burst arrival times showed more scatter relative to the observations than is typical for previous applications of this code, with an RMS error between the observed and predicted times of 0.85~hr. The residuals appear to be correlated, suggesting some of the differences may be attributed to unmodeled variations in the accretion rate.
The predicted $\alpha$-values are consistent with the observational values, although there is little range in the latter, and large uncertainties.
We also compare the observed peak fluxes with the expected (isotropic) Eddington limit at the source distance, and find that the later bursts are somewhat in excess of this value, by about 50\%. It is not uncommon to observe non-PRE bursts in excess of the peak flux of PRE bursts \cite[e.g.][]{minbar}, and the possible impact of the anisotropy of the PRE burst peak fluxes are not included, so we do not consider this discrepancy too serious.

\begin{figure*}[ht!]
    \begin{centering}
        \includegraphics[width=\linewidth]{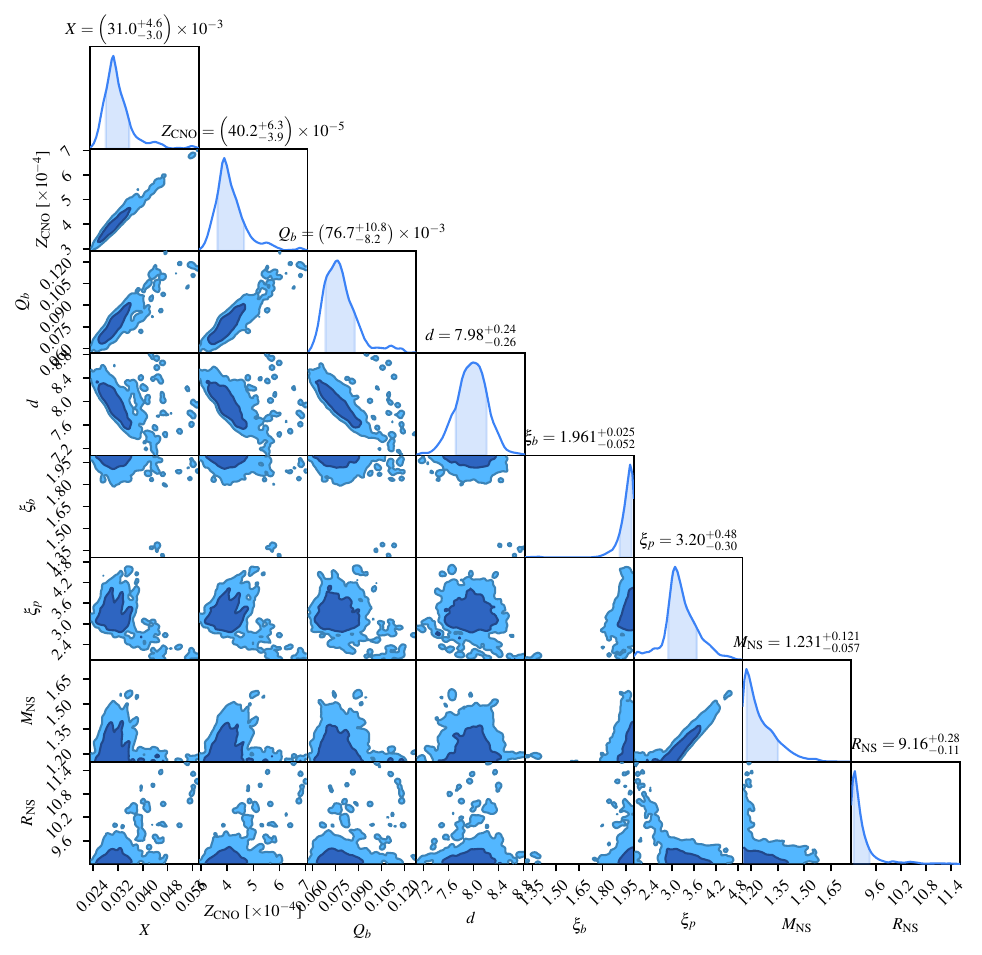}
        \caption{
            Model parameter posterior distributions for $5\times10^5$ samples of the simulation 
            to match the \sigr\ observation. 
            Contours are at 1 and $2\sigma$.
        }
        \label{fig:punkt_posteriors}
    \end{centering}
\end{figure*}

The posterior distributions are generally well constrained (Fig. \ref{fig:punkt_posteriors}), and most notably indicate remarkably low  H- and CNO-fractions for the burst fuel. $X_0$ is less than a tenth, and \zcno\ almost 1/30th of the expected solar values. As is common for these analyses, the posteriors exhibit a marked correlation between $X_0$ and \zcno.
We also observe correlations between other pairs of parameters, including \qb\ and $X_0$/\zcno; and distance $d$ and $X_0$/\zcno/\qb. The $\xi_b$ distribution is constrained by the prior upper limit of 2; conversely, both  the neutron star mass and radius are concentrated towards their lower limits, $1.15M\odot$ and 9~km, respectively.

\begin{figure}[ht!]
    \begin{centering}
        \includegraphics[width=\linewidth]{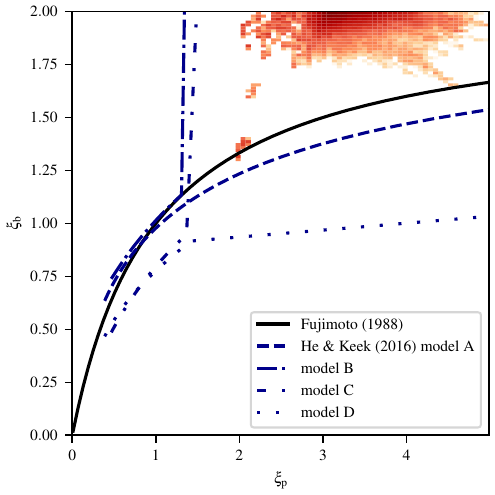}
        \caption{
        Distribution of inferred anisotropy parameters $\xi_b$, $\xi_p$ for the burst and persistent emission, respectively.
        The colour shading is based on the density of sample points within a grid, with darker red indicating higher density.
        The lines indicate different models for the anisotropy parameters as a function of inclination; for reference, low values of the anisotropy parameters correspond to low inclination (more face-on disk).
        }
        \label{fig:punkt_xis}
    \end{centering}
\end{figure}

The inferred distance of 8~kpc is somewhat in excess of the upper limit of 6.9~kpc derived from the peak burst fluxes by \cite{altamirano10b}, but perhaps of more concern is the combination with the large values for the anisotropy parameters $\xi_b$ and $\xi_p$, both well in excess of 1. These ranges indicate that both the burst and the persistent flux are beamed preferentially away from the line of sight, such that the inferred burst and persistent luminosities are much larger than would be estimated assuming isotropy. Another way to view this is that the inferred distance assuming isotropy of the emission is of order $d\sqrt{\xi}$, i.e. in the range 11--14~kpc. Such a large distance would translate to substantially super-Eddington bursts (given the measured peak fluxes), without any evidence for photospheric radius-expansion; as it is, the peak fluxes of the later bursts in the outburst are already slightly in excess of the inferred (isotropic) Eddington flux. While this beaming is fully consistent with the assumptions of our disk model,  the particular combination of parameters is not consistent with any specific model (Figure \ref{fig:punkt_xis}) and so must be viewed with some caution.

We undertook several different \ac{MCMC} runs to test how sensitive the model parameters were to the data treatment. The individual parameter ranges are listed in Table \ref{tab:igr17511_params}. 
The neutron star mass and radius 
tended to low values (constrained by the prior lower limits), so we undertook a run with fixed \mns, \rns, labelled {\tt pa}. The resulting posterior ranges were comparable, with slightly lower $X$ and larger $d$.
Given the issue with the large distance and hence possibly super-Eddington bursts, we introduced an additional contribution to the likelihood based on the peak fluxes, such that larger distances would be penalised. This addition reduced the distance slightly for fixed \mns, \rns\ (compare runs {\tt ppa} and {\tt pa} but not at all for the runs where the mass and radius were free ({\tt ppa\_mr} and {\tt pa\_mr}).
Finally, we made one run including both the peak flux term and the \mns, \rns\ prior of \cite{goodwin19c}, based in turn on the models of \cite{steiner18} derived from neutron stars in globular cluster. This run also produced parameter values comparable to the other runs, although with slightly higher mass and radius commensurate with the different prior ranges.

\begin{table*}
	\caption{System parameters for \sigr\ derived from the punctuated burst train simulations. 
    }
	\label{tab:igr17511_params}
\begin{tabular}{cccccccc}
\hline
Parameter & Units & Prior$^{{\rm a}}$ & {\tt pa} & {\tt pa\_mr} & {\tt ppa} & {\tt ppa\_mr} & {\tt ppa\_mrp}$^{{\rm b}}$ \\
\hline
\ensuremath{X} &  & $U[10^{-5},0.76]$ & $0.0228_{-0.0006}^{+0.0031}$ & $0.031_{-0.003}^{+0.005}$ & $0.0375_{-0.0046}^{+0.0015}$ & $0.031_{-0.003}^{+0.004}$ & $0.032_{-0.003}^{+0.002}$\\
\zcno & $10^{-4}$ & $U[10^{-5},0.056]$ & $3.28_{-0.09}^{+0.45}$ & $4.0_{-0.4}^{+0.6}$  & $4.7_{-0.6}^{+0.2}$ & $4.0\pm0.4$ & $4.1_{-0.4}^{+0.5}$ \\
\ensuremath{Q_b} & MeV/nucleon & $U[10^{-6},5]$ & $0.0438_{-0.0015}^{+0.0052}$ & $0.077_{-0.008}^{+0.011}$ & $0.066_{-0.009}^{+0.004}$ & $0.075\pm0.008$ & $0.072_{-0.007}^{+0.006}$\\
\ensuremath{d} & kpc & $U[1,20]$ & $9.60_{-0.32}^{+0.07}$ & $8.0_{-0.3}^{+0.2}$ & $8.71_{-0.10}^{+0.32}$ & $8.0\pm0.2$ & $8.16_{-0.18}^{+0.19}$ \\
\ensuremath{\xi_b} &  & $U[0.01,2]$ & $1.988_{-0.022}^{+0.009}$ & $1.97_{-0.05}^{+0.03}$ & $1.985_{-0.023}^{+0.011}$ & $1.97_{-0.05}^{+0.02}$ & $1.979_{-0.038}^{+0.016}$ \\
\ensuremath{\xi_p} &  & $U[0.01,10]$ & $3.00_{-0.05}^{+0.02}$ & $3.2_{-0.3}^{+0.5}$ & $2.87_{-0.04}^{+0.03}$ & $3.3_{-0.3}^{+0.5}$ & $3.3_{-0.4}^{+0.6}$\\
$M_{\rm NS}$ & $M_\odot$ & $U[1.15,2.5]$ & (fixed) & $1.22_{-0.06}^{+0.12}$ & (fixed) & $1.26_{-0.08}^{+0.14}$ & $1.30_{-0.12}^{+0.16}$\\
$R_{\rm NS}$ & km & $U[9,17]$ & (fixed) & $9.14_{-0.10}^{+0.27}$& (fixed) & $9.12_{-0.09}^{+0.20}$ & $9.59_{-0.07}^{+0.16}$ \\
$g$ & $10^{14}\ {\rm cm\,s^{-2}}$ & & & $2.50_{-0.19}^{+0.34}$ & & $2.6_{-0.2}^{+0.4}$ & $2.4_{-0.3}^{+0.4}$ \\
$1+z$ & & & & $1.29_{-0.02}^{+0.04}$ & & $1.30_{-0.03}^{+0.05}$ & $1.29_{-0.04}^{+0.06}$ \\
\hline
$N_{\rm steps}$ & & & $3\times10^4$ & $3\times10^4$ & $6\times10^4$ & $6\times10^4$ & $6\times10^4$\\
RMS error & hr & & 0.9381 & 0.8535 & 0.8563 & 0.8529 & 0.8550\\
$f_{\rm acc}$ & & & 0.3823 & 0.2815 & 0.3969 & 0.3173 & 0.3272 \\
{\tt stretch\_a} & & & 1.25 & 1.25 & 1.25 & 1.25 & 1.25 \\
\hline
\end{tabular}
  \begin{tabnote}
    {$^{{\rm a}}$The entries in the Prior column specify what prior functions are adopted, with $U(a,b)$ indicating a uniform distribution between $a$ and $b$.}\tnp
    {$^{{\rm b}}$For this run, we used instead the $M,R$ prior of \cite{goodwin19c}.}\tnp
  \end{tabnote}
\end{table*}

\subsection{Analytic constraints on fuel composition}
\label{sec:analytic}

Following the unexpectedly low composition values derived from the MCMC simulations, we revisited the burst properties to analytically verify the composition. The two key properties we highlight is the relatively constant burst fluence prior to the last-observed burst (varying only by $\approx20$\%, despite a drop in persistent flux (and hence accretion rate) of 50\%); and the much larger increase in fluence for the last burst. 

We consider two pairs of bursts: the second and third bursts (separated by $\Delta t_3=7.07$~hr), and the 
\#19 and 20 (separated by $\Delta t_{20}=15.77$~hr; Table \ref{tab:bursts}).
We used our spline-smoothed flux history (e.g. Fig. \ref{fig:outburst}) to calculate the average bolometric flux over both intervals $F_3$ and $F_{20}$, which were 1.71 and $0.842\times10^{-9}\ {\rm erg\,cm^{-2}\,s^{-1}}$, respectively. We can then calculate the ratio of accreted mass:
\begin{equation}
    \frac{\Delta M_{20}}{\Delta M_3} = \frac{F_{20}\Delta t_{20}}{F_3\Delta t_3} = 1.1
\end{equation}
i.e. 10\% more mass was accreted prior to ignition of the 20th burst, compared to the 3rd (assuming conservative burning; for the moment, we neglect the errors on the ratio).

The ratio of the fluences of the two bursts is then the product of the ratio of the accreted masses multiplied by the ratio of the nuclear burning energy $Q_{\rm nuc} \approx 1.35+6.05\overline{X}$, where $\overline{X}$ is the average H-fraction at ignition \citep{goodwin19a}:
\begin{equation}
    \frac{E_{b,20}}{E_{b,3}} = \frac{\Delta M_{20}}{\Delta M_3} \frac{Q_{{\rm nuc},20}}{Q_{{\rm nuc},3}} = 1.19 \pm 0.05
\end{equation}
We can invert this expression to solve for the ratio of nuclear burning energy for the two bursts:
\begin{equation}
    \frac{Q_{{\rm nuc},20}}{Q_{{\rm nuc},3}} =\frac{E_{b,20}}{E_{b,3}} \left( \frac{\Delta M_{20}}{\Delta M_3}\right)^{-1} =  1.08 \pm 0.04 
    \label{qnuc}
\end{equation}

The effect of H-burning is to reduce the average H-fraction $\overline{X}=X_0(1-0.5f_{\rm burn})$, where $X_0$ is the accreted composition,  $f_{\rm burn} = \Delta t/\Delta t_{\rm CNO}$ is the fraction of the H burnt at the base (for $\Delta t<\Delta t_{\rm CNO}$), and $t_{\rm CNO}$ is the time to exhaust the H at the base:
\begin{equation}
    t_{\rm CNO} = 9.8\left(\frac{X_0}{0.7}\right)\left(\frac{\zcno}{0.02}\right)\ {\rm hr}
    \label{tcno}
\end{equation}
\citep{concord22}.
The $Q_{\rm nuc}$ ratio between the 20th and 3rd burst intervals is  unexpectedly $>1$; $Q_{\rm nuc}$ depends linearly on the average H-fraction at ignition $\overline{X}$, and so should be {\it reduced} for longer recurrence times, as there is more time to burn the accreted hydrogen. (This discrepancy could arise from systematic errors affecting the averaged fluxes, for example). By that reasoning, we adopt only the part of the $Q_{\rm nuc}$ probability distribution that is $<1$, which is just outside the $2\sigma$ lower limit of the value.

We hypothesise that the more dramatic increase in fluence (accumulated mass) for the last burst is a consequence of the accreted hydrogen being exhausted at the base of the layer prior to that burst, whereas for the earlier bursts the hydrogen has not been exhausted \cite[cf. with][Figure 1]{gal06c}. That is, $t_{\rm CNO} \approx \Delta t_{20}$. With that assumption, we can substitute in equation \ref{tcno} to our expression for $\overline{X}$, substitute that into equation \ref{qnuc} and solve for $X_0$. 

\begin{figure}[ht!]
    \begin{centering}
        \includegraphics[width=\linewidth]{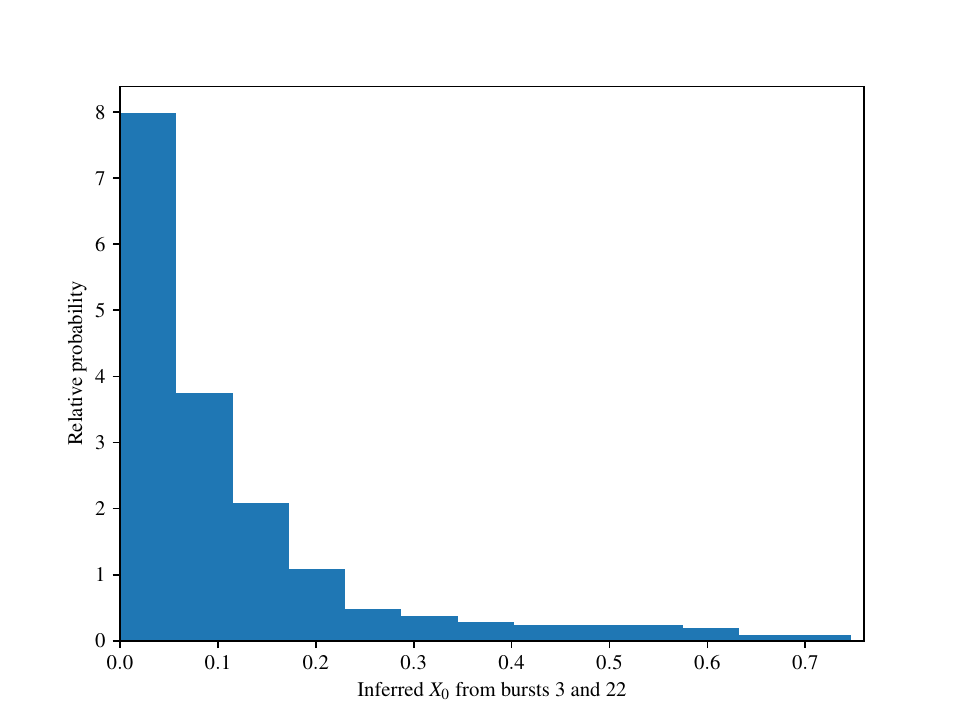}
        \caption{
        Inferred fuel H-fraction $X_0$ based on the properties of bursts \#3 and \#22, as described in the text. Note the very strong preference to small values, which qualitatively supports the outcomes of the {\sc beansp} simulations.
        }
        \label{fig:x0}
    \end{centering}
\end{figure}

Our derived distribution for $X_0$ (based on a Monte-Carlo calculation using the tail of the $Q_{\rm nuc}$ ratio in equation \ref{qnuc}) is shown in Fig. \ref{fig:x0}. This distribution is very strongly peaked around $X_0\approx0$, thus lending qualitative support to the MCMC results. And the low \zcno\ follows from our assumption for $t_{\rm CNO}$, which fixes the relation between $X_0$ and \zcno.

\subsection{Outburst history and long-term mass transfer}
\label{sec:outbursts}

Here we examine the long-term outburst history, with a view to constraining the long-term mass transfer rate and hence the distance.
In addition to the bolometric persistent flux measurements covering the 2009 outburst, we gathered persistent flux measurements covering the subsequent two outbursts. 

It is relevant here to ask how complete our coverage of the outbursts of \sigr\ is. 
If an intermediate outburst of the source in the period of interest was overlooked, the average X-ray flux calculated between outbursts will be underestimated. An outburst could be missed if it occurred when the region of sky near the Galactic centre was unobservable due to the proximity of the Sun; e.g. for the All-Sky Monitor (ASM) onboard \xte, such intervals can last for 30~d, enough to cover a full outburst of \sigr\ \cite[e.g.][]{wen06}. Outside those intervals, outbursts reaching at least a few by $10^{-10}\ \epcs$ should have been detectable from 1995 onwards \cite[]{heinke25}. The estimated peak 0.5--10~keV flux for the 2009 and 2015 outbursts were in the range 6--$8\times10^{-10}\ \epcs$, but the 2025 outburst reached a peak less than half these values. Thus, we expect it is likely that our sample is complete, but estimate a roughly 10\% chance of a missed intermediate outburst.

For each of the detected outbursts we generated a spline approximation to the persistent flux history, following the approach implemented in {\sc beansp} (see e.g. Fig. \ref{fig:outburst}). We then integrated the flux model over the approximate duration of the outburst to estimate the total fluence, listed in Table \ref{tab:outbursts}. As with the 2009 outburst, the variations to the persistent flux from measurement to measurement in the 2015 and 2025 outbursts was in excess of the uncertainties on the flux measurements. We conclude there are systematic errors in these outburst fluence estimates, that are difficult to quantify. Thus, we neglect the error on the fluence value for the 2009 outburst (this quantity is not so useful as we do not have a measured outburst interval) but propagate the uncertainty on the bolometric correction through to the fluences for the 2015 and 2025 outbursts. 

The fluences for the 2009 and 2015 outbursts were comparable, but the 2025 outburst appeared significantly shorter and weaker.  The relative weakness of the 2025 outburst is in contrast to the significantly longer wait time, of 9.9~yr (compared to 5.5~yr for the 2015 outburst) during which a significantly larger mass reservoir might have been expected to accumulate, if the mass transfer rate was constant.

Based on the recurrence time for the last two outbursts, we estimate a long-term average flux in the interval beforehand of 1.1 and $0.03\times10^{-10}\ {\rm erg\,cm^{-2}\,s^{-1}}$, respectively. For close binaries like \sigr, the long-term mass transfer rate $\dot{M}$ is expected to be driven by the loss of orbital energy from gravitational radiation \cite[e.g.][]{bc01}. Provided the mass transfer is conservative, and the companion mass and radius are related by $R_C\propto M_C^n$ where $n=-1/3$, we can equate the expected $\dot{M}$ with the long-term X-ray flux $\left<F_X\right>$ to obtain a lower limit on the distance:

\begin{eqnarray}
d& \gtrsim& 5.4        \left(\frac{\left<F_X\right>}{10^{-10}\ {\rm erg\,cm^{-2}\,s^{-1}}}\right)^{-1/2}
       \left(\frac{M_C}{0.1M_\odot}\right) \nonumber \\
     &  &\times\left(\frac{P_{\rm orb}}{2\ {\rm hr}}\right)^{-4/3}
       \left(\frac{M_{\rm NS}}{1.4M_\odot}\right)^{1/3}
       \left(\frac{1+z}{1.259}\right)^{-1/2} \nonumber \\
     &  & \times\left(\frac{Q_{\rm grav}}{193\ {\rm Mev\,nucleon^{-1}}}\right)^{1/2}
\xi_p^{-1/2}\ {\rm kpc}
\end{eqnarray}
This expression is equivalent to that given in \cite{gal06c}, but we have retained the dependence on $Q_{\rm grav}=cz/(1+z)$ (the accretion yield) and redshift $1+z$, and also included the $\xi_p$ term parametrising the possible anisotropy of the persistent emission.
Substituting in the values for \sigr, with $\left<F_X\right>$ from Table \ref{tab:outbursts}; minimum companion mass $M_C=0.134M_\odot$ \cite[for $M_{\rm NS}=1.4M_\odot$][]{papitto10}; $P_{\rm orb}=3.47$~hr; and standard values for the neutron star mass and radius (and hence $Q_{\rm grav}$ and $1+z$), we obtain $d\xi_p^{1/2}\gtrsim10.5$~kpc for the 2009--15 outburst interval, and $d\xi_p^{1/2}\gtrsim20$~kpc for the 2015--25 outburst interval.
From the posterior distributions of our preferred run {\tt pa\_mr}, $d\xi_p^{1/2} = 14.3_{-0.9}^{+1.1}$~kpc.
Thus, while the former limit is consistent with the distance inferred from our burst analysis with the posterior distribution of $\xi_p$ that we also obtain, the latter is not. 

This discrepancy implies that we are detecting insufficient flux given the expected rate of mass transfer, and suggests that either the mass transfer is not conservative, or that the radius of the companion responds differently to the mass transfer than would a main sequence donor.

\section{Discussion}
\label{sec:discussion}

We compared the thermonuclear bursts observed from \sigr\/ during its 2009 outburst, with the predictions of a numerical ignition model. We found the best agreement for low values of the  hydrogen and CNO mass fractions, with $X\approx0.03$ and $\zcno \approx 4\times10^{-4}$, suggesting an unexpectedly helium-rich and metal-poor donor, at a distance that is difficult to reconcile with the measured burst peak fluxes and the lack of observed photospheric radius-expansion.

The simultaneously low values of $X$ and \zcno\ may have implications for the role of sedimentation and spallation in the accreted material. Sedimentation will tend to concentrate heavy elements in deeper layers, at the expense of lighter \cite[e.g.][]{pbt07}, which is not what we infer for \sigr. On the other hand, spallation may instead destroy heavy elements as they are accreted \cite[e.g.][]{randhawa19}, which might explain the low \zcno\ value, but not why this process operates in \sigr\ but not others.
Our previous analyses of \ssax, IGR~J17498$-$2921 \cite[]{gal24} and SRGA~J144459.2$-$604207 (\citealt{gal26}; see also \citealt{dohi25}) indicated slightly H-deficient donors, and such deficiencies have been attributed to the effects of long-term mass loss over the evolutionary history of the system \cite[e.g.][]{goodwin20a}. However the \zcno\ values for those sources were consistently in the range 0.01--0.02, broadly compatible with the solar value.
The diversity of compositional parameters between the sources analysed so far by this method instead suggests that the compositions are intrinsic and are only weakly affected by either sedimentation or spallation, at least in the range of accretion rates explored with these data.

We also examined the long-term accretion outburst history, and estimated the long-term average X-ray flux. Our derived distance from the burst analysis is incompatible with the lower limits derived from the expected mass transfer rate predicted from the gravitational radiation of the system. 

Our analytic approach for this source adopts a number of new features to our burst comparison method, including decoupling the successive bursts in the train to reduce the cumulative effect of model errors. 
We anticipate that these improvements will make the code and algorithm more robust and widely applicable in future.
At the same time, the lack of formal convergence of the method remains a concern, and future work is planned to investigate the possible reasons, including by using different MCMC samplers.

A few other binary systems may share similar properties to \sigr.
The well-known persistent burst source 4U~1728$-$34 \cite[][]{hoff76} was long thought to be an ultracompact system thanks to its consistently short bursts and X-ray flux variations \cite[see e.g.][]{gal10b}, but more recently measurements of lags between bursts and reprocessed IR radiation have shown that the orbital period $\gtrsim3$~hr \cite[]{vincentelli20,vincentelli23}. Additionally, a binary system with a 3.6~hr orbit consisting of a mildly recycled pulsar and a He star companion with mass 1.0--$1.6\ M_\odot$ has recently been discovered \cite[]{yang25}. Based on the X-ray mass function, such a massive donor could not be present in \sigr\ unless the inclination was implausibly low. However, a lower mass He donor could be present, with the only remaining caveat being that it would not fill its Roche lobe (adopting the mass-radius relations of \citealt{yang25}). 

The inferred system parameters for \sigr\ thus presents several challenges to fully explaining the properties and origin of the system.

\subsection{The source distance}

The inferred distance range of 8--9~kpc (depending on the specific {\sc beansp} run; Table \ref{tab:igr17511_params}) is in excess of the most conservative upper limit of 7.1~kpc derived from the peak flux of the bursts \cite[]{altamirano10b}, although this limit is calculated assuming isotropy of the burst emission. 
Our comparison of the observed peak fluxes with the (isotropic) maximum flux derived from the expected Eddington luminosity of $3.79\times10^{38}\ \eps$ \cite[]{kuul03a} indicates that these bursts would be slightly above that limit. On the other hand, the impact of the substantial burst anisotropy factor $\xi_b$ on the average peak flux of the bursts observed with \xte/PCA and \igr/JEM-X of $(43.9 \pm 7.4)\times10^{-9}\ \epcs$ \cite[]{falanga11} suggests that the intrinsic luminosity might be around $7\times10^{38}\ \eps$, well in excess of the limit.

It is possible that the last and likely most energetic burst during the 2009 outburst, observed with {\it Swift} \cite[]{bozzo10a}, could have exhibited radius expansion; but the first 5~s of this event were not observed with the XRT and so the peak flux (which is still within the range of the bursts observed by the other instruments) must be considered a lower limit.

Thus, a residual tension with our {\sc beansp}-inferred system parameters is that the bursts may substantially exceed the empirical Eddington limit, without exhibiting any indications of photospheric radius-expansion.

\subsection{The mass donor and accretion channel}
\label{sec:donor}

The low $X_0$ and \zcno\ values are substantially different from the expectations of the main-sequence donor suggested to be present in this system \cite[e.g.][]{papitto10}. We are  confident with the qualitatively low values for these parameters, as they can be deduced independently of the burst model-observation comparison with {\sc beansp}, by simple arguments related to the expected rate of steady H-burning (see Section \ref{sec:analytic}). 

Such low but non-zero hydrogen fractions only exist in regions of stars that have undergone substantial hydrogen burning. They can be found (i) near the centre of low-mass ($M_\mathrm{ZAMS}\lesssim1.2~\msun$) stars approaching the end of the main sequence, (ii) inside or just outside the convective core of higher-mass ($M_\mathrm{ZAMS}\gtrsim1.5~\msun$) stars near the end of the main sequence, or (iii) outside the helium core of early post-main-sequence stars. The inferred surface hydrogen abundance suggests that the donor of \sigr\ has undergone substantial mass loss via binary interactions to expose these deeper layers. The first possibility is disfavoured by the lower limit on the present donor mass, $M\gtrsim0.14~\msun$, constrained from the observed mass function and the absence of eclipses \citep{papitto10}. In low-mass stars, where hydrogen burning proceeds through the pp-chain, such low hydrogen abundances are confined to a relatively small central region. Exposing material with $X\sim\xlo$ would therefore require stripping the star to $M\lesssim0.1M_\mathrm{ZAMS}$, below the observational constraint.

The second scenario may have difficulties explaining the current orbital period. The convective cores of massive main-sequence stars often occupy $\sim1/5$--$1/2$ of the total stellar mass, and the fraction shrinks as the central hydrogen abundance declines \citep{shi25}. To expose material with $X\sim\xlo$, the donor must have developed a hydrogen-deficient core larger than the size of the current stellar mass before undergoing mass transfer. This requires mass transfer to begin relatively late in the main sequence, when the donor has already expanded considerably. The binary must then undergo sufficient orbital contraction to reach the observed period of $3.5~\mathrm{hr}$ while stripping the donor down to the hydrogen-deficient layers. Although isotropic re-emission can produce substantial orbital shrinkage for sufficiently large initial mass ratios, the orbit eventually re-expands as the donor loses mass. The parameter space satisfying both the required stripping depth and orbital contraction is therefore likely to be restricted. Moreover, such extreme mass ratios raise the question of whether mass transfer can remain dynamically stable throughout this evolution without merging.

This leaves us with only the third option: a post-main-sequence star stripped down to the layer immediately outside the helium core. A possible evolutionary pathway is case B mass transfer from an initially more massive ($M_\mathrm{ZAMS}\gtrsim1.5~\msun$) star. Following core hydrogen exhaustion, the contraction of the helium core drives the expansion of the stellar envelope. As the star evolves across the Hertzsprung gap, it may fill its Roche lobe and initiate mass transfer onto the neutron-star companion. Depending on the donor stellar structure and the mass ratio, the mass transfer process may become dynamically unstable, leading to a common-envelope phase. The subsequent inspiral can eject most of the hydrogen-rich envelope, leaving behind a stripped star with a thin, hydrogen-deficient envelope above its helium core.

\begin{figure}[ht!]
     \begin{centering}
         \includegraphics[width=\linewidth]{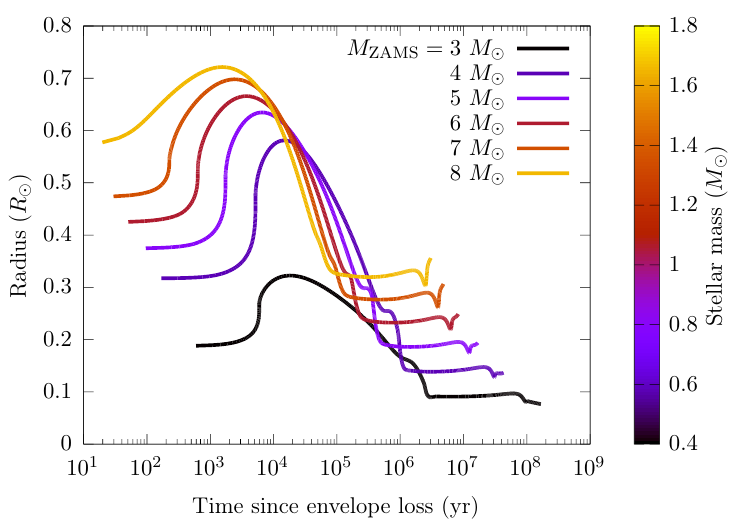}
         \caption{Radius evolution of our partially stripped giant models computed with {\sc MESA}. Time is measured since envelope loss, which is assumed to be during the Hertzsprung gap phase. The curves are coloured based on their post-stripping stellar mass, with the corresponding initial masses indicated in the legend.
         }
         \label{fig:radius_evolution}
     \end{centering}
 \end{figure}

The current donor can thus be considered as an intermediate object between a conventional hydrogen-rich main-sequence star and a pure helium star. Our {\sc mesa} simulations (Section \ref{sec:mesa}) demonstrate that the mass--radius relation of such stars can be substantially different from either sequence. 
Figure~\ref{fig:radius_evolution} shows examples of how the radius evolves after mass stripping. The star expands immediately following the sudden mass loss, and then subsequently contracts over a timescale of $\sim10^{5\mathrm{-}6}$~yr as it thermally re-adjusts to the new mass. The star continues to slowly contract beyond shell hydrogen ignition (the envelope mass is not sufficient to support giant-like structures), after which the helium core starts to grow at the expense of the hydrogen-poor envelope mass. Following core helium ignition, the star settles into a phase of relatively stable radii that persists until core helium exhaustion.

 \begin{figure}[ht!]
     \begin{centering}
         \includegraphics[width=\linewidth]{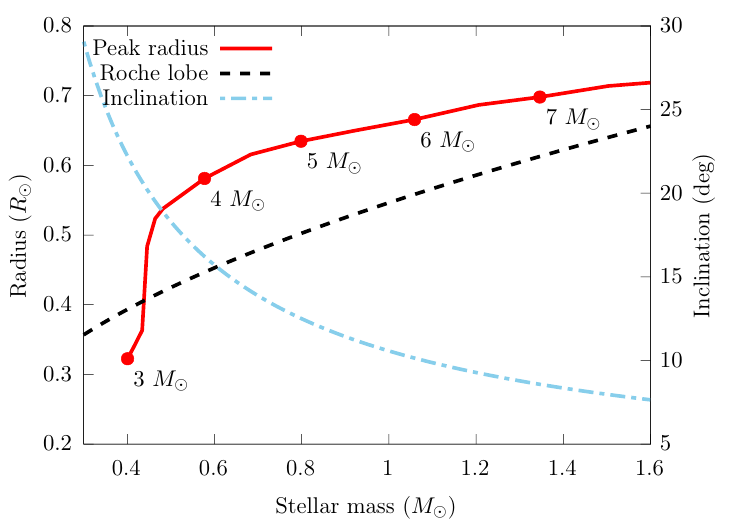}
         \caption{Maximum radii of our partially-stripped giant {\sc mesa} models as a function of stellar mass (solid red curve). We label the curve with corresponding initial masses $M_\mathrm{ZAMS}$. The dashed black curve shows the donor Roche lobe radius for the given mass assuming $M_\mathrm{NS}=1.4~\msun$ and an orbital period $P_\mathrm{orb}=3.5~\mathrm{hr}$. The dot-dashed blue curve shows the corresponding orbital inclination based on the observed mass function \citep{riggio11}.
         }
         \label{fig:M-R_relation}
     \end{centering}
 \end{figure}

 The stellar radius is sensitive to the thermal profile and can strongly depend on the details of the mass transfer process, which are beyond the scope of this exploratory study. Nevertheless, our models provide a useful guideline as to which stellar masses can potentially overflow its Roche lobe. In Figure~\ref{fig:M-R_relation}, we plot the maximum radii achieved in our {\sc mesa} models as a function of post-stripped stellar mass along with the corresponding Roche-lobe radii at the observed orbital period of \sigr\ ($P_\mathrm{orb}\sim3.5$~hr). We find that models with post-stripping masses $M\gtrsim0.45~\msun$ ($M_\mathrm{ZAMS}\gtrsim3.3~\msun$) attain radii exceeding their Roche-lobe radii, satisfying the necessary condition for Roche-lobe overflow in the observed system. The radii of these partially stripped stars generally lie between those of pure helium stars and normal main-sequence stars of comparable masses, illustrating the importance of the residual helium-enriched envelope.

Combining the above mass constraint with the observed binary mass function implies an orbital inclination of $i\lesssim20~\mathrm{deg}$, assuming $M_\mathrm{NS}=1.4~\msun$ (Figure~\ref{fig:M-R_relation}). We note that this limit is in tension with the implied high inclination from the large values of $\xi_p$, $\xi_b$ from the {\sc beansp} simulations, as well as the constraints suggested from spectral fitting of the iron K$\alpha$ line \cite{papitto10}, of 38--68$^\circ$. Given the preliminary nature of our stellar modelling, the lack of verification of the disk anisotropy models, and possible systematic errors in the iron line fitting, we do not consider this discrepancy as critical at this point, but acknowledge that more detailed work is needed.

Many aspects of common-envelope evolution remain poorly understood, including the conditions for its onset, the final orbital separation, and the amount of envelope mass retained by the donor. In particular, the mass coordinate separating the ejected envelope from the bound stellar remnant remains a point of debate. Various criteria have been proposed to predict this bifurcation point, based on quantities such as the nuclear generation rate \citep{tau01,dem11}, hydrogen fraction \citep{dew00,xu10} and thermodynamic quantities \citep{han94,iva11}. If \sigr\ is indeed a post-common-envelope binary, the inferred surface hydrogen abundance of $X\sim\xlo$ may provide an empirical lower limit on the depth of envelope ejection. Furthermore, the ongoing mass transfer may be consistent with the extended, thermal-timescale mass-transfer phase proposed to follow the initial dynamical inspiral \citep{mey79,hir22}. Together with the present orbital period, these properties offer potential constraints on both the stripping depth and the orbital evolution during and after the common-envelope phase.

 \subsection{Magnetically channelled accretion flow?}

A possible resolution to the various puzzling properties of the \sigr\ system is suggested by the observation that the pulse fraction, at around 20\%, is the highest of all the accretion-powered millisecond pulsars, and consequently the harmonic content is complex, with up to a fourth harmonic present sporadically \cite[]{papitto11,riggio11}.
We speculate that the high pulse fraction and complex profile is a consequence of magnetically channelled flow, occurring to a greater extent than in any other bursting \acp{AMP}. The flow is more strongly channelled because the magnetic field is stronger, as the system is younger than other such objects, roughly the main-sequence lifetime of the mass donor, $\lesssim 250$~Myr. Thus, the magnetic field has had less time to decay down to the $\sim10^8$--$10^9$~G levels inferred for the majority of \acp{AMP} \cite[e.g.][]{disalvo22}.

The channelled flow focuses the accretion onto a fraction $f$ of the neutron star area, resulting in higher local accretion rate $\dot{m}=\dot{M}/4\pi R^2f$ than if the accretion was over the entire surface. Because our burst comparison seeks to effectively match the $\dot{M}$ inferred from the persistent flux with the burst arrival times that depend instead on $\dot{m}$, channelled flow is expected to 
bias the {\sc beansp} algorithm to distances larger than the true value. Similarly, since the observed burst fluence $E_b=E_{\rm tot}/4\pi d^2\xi_b$, a smaller $E_{\rm tot}$ (resulting from a smaller fuel reservoir) will also 
bias the distance in the same direction. Ignition of fuel over only a partial region of the neutron star surface may also help to reconcile the unusually large values of the anisotropy parameter $\xi_b$ (if not also $\xi_p$) derived from the burst comparison.
Future modifications to the {\sc beansp} code may make such a scenario testable.

\begin{acknowledgements}
We thank Alessandro Papitto and Giulia Illiano for providing the flux measurements covering the 2015 and 2025 outbursts of \sigr.
DKG acknowledges the support of the International Space Science Institute (ISSI) in Bern, Switzerland, during two periods as visiting scientist in late 2024 and early 2025.
AJG is grateful for support from the Forrest Research Foundation. 
RH acknowledges support from the RIKEN Special Postdoctoral Researcher Program for junior scientists.
We also acknowledge the input, support and encouragement of the late Prof. Dr. Maurizio Falanga (ISSI/Univ. Bern).
\end{acknowledgements}

\section*{Data Availability}

The \sigr\ burst data (Table \ref{tab:bursts}) is available as a machine-readable table accompanying this paper. The version used for input to  {\sc beansp} is available at the Github repository\footnote{\url{https://github.com/outs1der/beans}}.
A snapshot of the code and the posterior samples from the last $10^3$ steps of each of the runs presented here can be found at the 
Monash Bridges repository\footnote{\url{https://doi.org/10.26180/34994695}}.

\bibliography{all,bib}

\end{document}